\documentclass{article} \usepackage{epsf} \usepackage{amsmath}
\usepackage{bbm}

\newcommand{\be}{\begin{equation}} 
\newcommand{\ee}{\end{equation}}

\newcommand{\half}{\frac{1}{2}}

\newcommand{\bPLR}{\bar{P}_{\mp}}
\newcommand{\bPRL}{\bar{P}_{\pm}}

\newcommand{\PLR}{{P}_{\mp}}
\newcommand{\PRL}{{P}_{\pm}}

\newcommand{\Gammaf}{\Gamma_5} \newcommand{\bGammaf}{\overline{\Gamma}_5}

\newcommand{\One}{\ensuremath{\mathbbm 1}}

\newcommand{\N}{\mathcal{N}}

\newcommand{\psibar}{\overline{\psi}}

\newcommand{\T}{\mathcal T}
\newcommand{\Pa}{\mathcal P}

\newcommand{\spro}[2]{(#1\, , \,#2)}
\newcommand{\sproO}[3]{\spro{#1}{#2\,#3}}
\newcommand{\npoint}[1]{\sproO{\Omega}{#1}{\Omega}}

\newcommand{\com}[2]{[#1,#2]}

\begin{document}
\thispagestyle{empty} \parskip=12pt \raggedbottom
 
\def\mytoday#1{{ } \ifcase\month \or January\or February\or March\or
  April\or May\or June\or July\or August\or September\or October\or
  November\or December\fi
%\space\number\day , \\
%\space \number\time , \\
  \space \number\year}
\noindent
%\hspace*{6cm} BUTP/\\
\vspace*{1cm}
\begin{center}
  {\LARGE $CP$, $T$ and $CPT$ in the non-perturbative formulation of chiral
    gauge theories}

%\footnote{Work supported in part by the Schweizerischer Nationalfonds}
 
  \vspace{1cm} Peter Hasenfratz, Moritz Bissegger\\
  Institute for Theoretical Physics \\
  University of Bern \\
  Sidlerstrasse 5, CH-3012 Bern, Switzerland
  
  \vspace{0.5cm}
  
  \nopagebreak[4]
 
\begin{abstract}
In spite of significant recent progress on the non-perturbative formulation of chiral
gauge theories there remained several unsolved problems. One of them is the
puzzle that the left- and right-handed projectors, and so the left- and right-handed actions, break $CP$-symmetry on the lattice. We show in this letter
that they break $T$-symmetry also, while $CPT$ remains intact.
\end{abstract}

\end{center}
\eject

\subsubsection*{Introduction}
Chiral symmetry is a longstanding problem on the lattice. Under very general
conditions, the chiral transformations defined by simple $\gamma^5$ rotations
on the fermions fields can not leave the lattice action invariant
\cite{NN1981}. This deadlock was resolved in a rather unexpected way: lattice
formulations with a Dirac operator $D$ satisfying the Ginsparg-Wilson relation \cite{GW1982}
\be
D\gamma^5+\gamma^5D=D\gamma^5D\label{GWr}
\ee
have an exact symmetry transformation which depends on the Dirac operator
itself and can be considered as a realization of the formal continuum
transformation \cite{L1998, summary}. The projectors to left- and right-handed fermions
also depend on $D$. Actually, the form of the projectors is not unique. One might
introduce a continuous parameter $s$ and for each value of $s$ the fermion
action is decomposed correctly
\be
\psibar D\psi=\psibar^{(s)}_L D\psi^{(s)}_L+\psibar^{(s)}_R D\psi^{(s)}_R.\label{chiralfrag}
\ee 
where
\begin{align}
\psi^{(s)}_{L/R}&=\PLR^{(s)}\psi,&\psibar^{(s)}_{L/R}&=\psibar\bPRL^{(s)}.
\end{align}
and the projectors are defined as
\begin{align}
\PLR^{(s)}&=\frac{1}{2}\left(\One\mp\Gammaf^{(s)}\right),&\bPLR&=\frac{1}{2}\left(\One\mp\bGammaf^{(s)}\right),\nonumber\\
\Gammaf^{(s)}&=\left(\N^{(s)}\right)^{-1}\gamma^5\left(\One-sD\right),&\bGammaf^{(s)}&=\left(\N^{(s)}\right)^{-1}\left(\One-(1-s)D\right)\gamma^5,\label{projectors}\\
\N^{(s)}&=\sqrt{\One-s(1-s)DD^\dagger}\nonumber.
\end{align}
Beyond being gauge field dependent, the projectors in eq.(\ref{projectors})
are asymmetric with respect to $\psi$ and $\psibar$. One might interpret this
asymmetry as a signal for a possible fermion number anomaly in chiral gauge
theories \cite{thooft}. Indeed, using the index theorem \cite{HLN}, it is easy
to connect the fermion number violation to the topological charge of the gauge
field configurations. On the other hand, the decomposition in
eq.(\ref{chiralfrag}) violates $CP$-symmetry also. Under the $CP$
transformation\footnote{In our convention \mbox{$W\doteq\gamma^2$},
  $\Pa n\doteq (-n^1,-n^2,-n^3,n^4)$ and the
  transposition $T$ , as well as the hermitian \mbox{conjugation $\dagger$},
  act always only on the Dirac and
  gauge indices.}
\begin{align*}
\psi(n)&\to W\psibar^T(\Pa
n),&\psibar(n)&\to-\psi^T(\Pa n)
W^{-1},
\end{align*}
\be
U_{\mu}(n)\to U_{\mu}^{(CP)}(n)\doteq\begin{cases}
  \left[U_{\mu}(\Pa n-\mu)\right]^T&(\mu=1,2,3)\\
 U^*_{\mu}(\Pa n)&(\mu=4)\end{cases}\label{CP}
\ee
and with the $CP$-symmetry condition for the Dirac operator
\be
W^{-1}D(n,n';U^{(CP)})^T W\doteq D(\Pa n',\Pa n;U)
\ee
 the left-handed action in eq.(\ref{chiralfrag}) is not invariant \cite{Hasenfratz:2001bz}
\be
\sum_{n,n'}\psibar^{(s)}_L(n) D(n,n';U)\psi^{(s)}_L(n')\stackrel{CP}{\to}\sum_{n,n'}\psibar^{(1-s)}_L(n) D(n,n';U)\psi^{(1-s)}_L(n').\label{CPleftaction}
\ee
At \mbox{$s=\half$} the $CP$-symmetry is restored, but just at this value of $s$ the
projectors become non-local: the $\lambda=2$ point, where the normalization
operator $\N^{(\half)}$ becomes zero, is an accumulation point of the
ultraviolet modes. The significance of this $CP$-symmetry violation on
the physical properties of the theory is an interesting, but not yet
completely understood issue \cite{Fujikawa:2002vj}.\\
We make here a short remark only on this difficult problem. Due to a technical
constraint in the treatment of chiral gauge theories on the lattice the theory
falls into topological sectors where the relative weight factors remain
undetermined. A theory defined on the Gaussian fixed-point has a finite number
of free parameters only and so these complex overall factors weighting the
partition function in the different topological sectors can not be simply {\it
  chosen} to satisfy the requirement of symmetry. Actually, the topological charge \mbox{$Q=0$}
sector should give informations on the \mbox{$Q\neq 0$} sectors due to the
clustering property of separated topological objects.\\
In this letter we investigate a simpler problem, namely the properties of the left- and right-handed
action under time reflection $T$ and $CPT$.

\subsubsection*{Time reflection in Euclidean space}
Time reflection is represented by an anti-unitary operator in Minkowski space
and this feature has its trace in Euclidean space also. We summarize below briefly
the transformation of fermion and gauge fields under time reflection in Euclidean path integral
formulation.\\
In Minkowski space the fermion operators are transformed as\linebreak
\mbox{$\psi(x)\to (-\gamma^1\gamma^3)\psi(\T x)$},
\mbox{$\psibar(x)\to \psibar(\T x)\gamma^1\gamma^3$} and
\mbox{$\T x\doteq(-x^0,x^1,x^2,x^3)$} \cite{Weinberg}. Having in mind
the path integral formulation it is useful to express the time reflection
symmetry on Green's functions
\begin{multline}
\npoint{T(\psi_{\alpha_n}(x_n)\cdot...\cdot\psi_{\alpha_1}(x_1)\psibar_{\beta_{m}}(\bar{x}_{m})\cdot...\cdot\psibar_{\beta_{1}}(\bar{x}_{1}))}\\
=\begin{aligned}[t](\Omega,\,T(&[-\gamma^1\gamma^3\psi(\T
  x_n)]_{\alpha_n}\cdot...\cdot[-\gamma^1\gamma^3\psi(\T
  x_1)]_{\alpha_1}\\
                   &\times[\psibar(\T\bar{x}_{m})\gamma^1\gamma^3]_{\beta_{m}}\cdot...\cdot[\psibar(\T\bar{x}_{1})\gamma^1\gamma^3]_{\beta_{1}})\Omega)^*,
\end{aligned}\label{Tsymmgfctstar}
\end{multline} 
where the complex conjugation of the expectation value on the right hand side is a consequence of the
anti-unitarity of the time reflection operator. Eq.(\ref{Tsymmgfctstar}) can also be
written as
\begin{multline}
\npoint{T(\psi_{\alpha_n}(x_n)\cdot...\cdot\psi_{\alpha_1}(x_1)\psibar_{\beta_{m}}(\bar{x}_{m})\cdot...\cdot\psibar_{\beta_{1}}(\bar{x}_{1}))}\\
=\begin{aligned}[t](\Omega\,&
,T([\gamma^3\gamma^1\gamma^0\psi(\T
\bar{x}_1)]_{\beta_1}\cdot...\cdot[\gamma^3\gamma^1\gamma^0\psi(\T
\bar{x}_m)]_{\beta_m}\\
&\times[\psibar(\T
  x_1)(-\gamma^0\gamma^3\gamma^1)]_{\alpha_1}\cdot...\cdot[\psibar(\T
  x_n)(-\gamma^0\gamma^3\gamma^1)]_{\alpha_n})\,\Omega).\end{aligned}\label{Tsymmgfct}
\end{multline}
On a Euclidean lattice these Green's functions are represented by integrals
over the Grassmann valued fields $\psi(n)$ and $\psibar(n)$. The symmetry
transformation on these fields corresponds to introducing new integration
variables which makes the left and right hand sides of eq.(\ref{Tsymmgfct})
identical. These variable substitutions read
\begin{align}
\psi(n)&\to-A^{-1}\psibar^T(\T
n),&\psibar(n)&\to \psi^T(\T
n)A,&
A&=-\gamma^1\gamma^3\gamma^4,\label{Tfields}
\end{align}
where the matrix $A$ has the property
\be
-A=A^{-1}=A^T=A^\dagger=\gamma^5A\gamma^5\label{Aproperties}
\ee
and \mbox{$\T n\doteq(n^1,n^2,n^3,-n^4)$}. The transformation in eq.(\ref{Tfields}) reflects the 4th axes, rotates the
Dirac indices and makes a swap between $\psi$ and $\psibar$. This last feature
is a consequence of the anti-unitarity of the time reflection operator.\\
Considering fermions in interaction with gauge fields the parallel transporter
$U_\mu(n)$, associated with the link $(n,n+\mu)$ of the lattice, enters. A simple
way to find the transformation law of $U_\mu(n)$ under time reflection is as
follows. Consider a free fermion theory on the lattice which is symmetric
under the transformation in eq.(\ref{Tfields}) (the standard free Wilson
action for example). Then introduce parallel transporters $U_{\mu}(n)$ to
make it gauge invariant and demand time reflection symmetry for this
interacting theory. One obtains
\be
U_{\mu}(n)\to U^{(T)}_{\mu}(n)\doteq\begin{cases} U^*_\mu(\T
  n)&(\mu=1,2,3)\\
[U_{\mu}(\T n-\mu)]^T&(\mu=4)
\end{cases}.\label{Tlink}
\ee
Consider now a general vector gauge theory on the lattice
\be
\sum\psibar_{\alpha}^a(n)D(n,n';U)_{\alpha\beta}^{ab}\psi_{\beta}^b(n'),\nonumber
\ee
where the sum runs over the space-time ($n$, $n'$), gauge ($a$, $b$) and Dirac
($\alpha$, $\beta$) indices. Performing time reflection using the
transformation rules in eqs.(\ref{Tfields}, \ref{Tlink}) and demanding
symmetry leads to the constraint on the Dirac operator $D$:
\be
AD(n,n';U^{(T)})^TA^{-1}\doteq D(\T n',\T n;U)\label{TsymmDirac}.
\ee

\subsubsection*{Time reflection for the left-handed action}
We shall assume that the Dirac operator $D$ satisfies the Ginsparg-Wilson
relation eq.(\ref{GWr}), has the standard hermiticity property
\be
D(n,n';U)=\gamma^5D(n',n;U)^\dagger\gamma^5\label{Dhermiticity},
\ee 
and obeys eq.(\ref{TsymmDirac}) which is the condition of time reflection
symmetry in a vector theory. It follows then
\begin{align}
\com{D}{D^\dagger}&=0,&\com{\gamma^5}{DD^\dagger}=0\label{comD}.
\end{align}
Using eqs.(\ref{projectors}, \ref{TsymmDirac}, \ref{Dhermiticity}, \ref{comD}) and the
properties of the matrix $A$ in eq.(\ref{Aproperties}) it is easy to show that
\begin{eqnarray}
A\left[\Gammaf^{(s)}(n,m;U^{(T)})\right]^TA^{-1}&=&-\bGammaf^{(1-s)}(\T m,\T n;U),\label{TGammafT}\\
A\left[\bGammaf^{(s)}(n,m;U^{(T)})\right]^TA^{-1}&=&-\Gammaf^{(1-s)}(\T m,\T n;U)\label{TbGammafT},
\end{eqnarray}
leading to
\begin{eqnarray}
A[\PLR^{(s)}(n,m;U^{(T)})]^TA^{-1}&=&\bPRL^{(1-s)}(\T m,\T n;U),\label{TPLRT}\\
A[\bPLR^{(s)}(n,m;U^{(T)})]^TA^{-1}&=&\PRL^{(1-s)}(\T m,\T n;U).\label{TbPLRT}
\end{eqnarray}
Eqs.(\ref{TPLRT}, \ref{TbPLRT}) imply that the left-handed action violates
time reflection in a similar way as $CP$ did
\be
\sum\psibar^{(s)}_{L}(n)D(n,n';U)\psi^{(s)}_{L}(n')\stackrel{T}{\to}\sum\psibar^{(1-s)}_{L}(n)D(n,n';U)\psi^{(1-s)}_{L}(n').\label{Tleftaction}
\ee
On the other hand, $CPT$ is a symmetry of the left-handed action independent
of the value of the parameter $s$, since both transformations swap $s$ and
$(1-s)$.\\ 
Combining the $CP$ and $T$ transformations of eqs.(\ref{CP},
\ref{Tfields}, \ref{Tlink}) it is easy to see that the $CPT$
transformation is a hypercubic rotation on the lattice
\begin{align}
\psi(n)&\to \gamma_5\psi(-n),&\psibar(n)&\to \psibar(-n)\gamma_5,&U_{\mu}(n)&\to
U_\mu^\dagger(-n-\mu).
\end{align}
Hypercubic symmetry implies then invariance under $CPT$.\\
 
{\bf Acknowledgements}
The authors thank Reto von Allmen for discussions.
This work was supported by the Schweizerischer Nationalfonds.

\eject

%%%\newpage

\end{document}